\documentclass[onecolumn,10pt]{article}
\usepackage{amsmath}
\usepackage{amsfonts}
\usepackage{amssymb}
\usepackage{amsthm}
\usepackage{times}
\usepackage{cancel}
\usepackage{dsfont}
\usepackage{graphics}

\newcommand{\trace}[1]{\mbox{$\mathrm{Tr}$}(#1)}

\newcommand{\ket}[1]{| #1 \rangle}
\newcommand{\bra}[1]{\langle #1 |}

\begin{document}

\date{}
\title{\textbf{`Two Dogmas' Redux}}
\author{\textbf{Jeffrey Bub} \\
\small Department of Philosophy\\
\small Institute for Physical Science and Technology\\\small Joint Center for Quantum Information and Computer Science\\
 \small University of Maryland, College
Park, MD 20742\thanks{\textit{Email address:} jbub@umd.edu} }
\maketitle

\section{Introduction}
About ten years ago, Itamar Pitowsky and I wrote a paper,  `Two dogmas about quantum mechanics' \cite{Bub+}, in which we outlined an information-theoretic interpretation of quantum mechanics as an alternative to the Everett interpretation.  Here I revisit the paper and, following Frauchiger and Renner \cite{Frauchiger+}, I show that the Everett interpretation leads to modal contradictions in `Wigner's-Friend'-type scenarios that involve `encapsulated' measurements, where a super-observer  (which could be a quantum automaton), with  unrestricted ability to measure any arbitrary observable of a complex quantum system, measures the memory of an observer system (also possibly a quantum automaton) after that system measures the spin of a qubit. In this sense, the Everett interpretation is inconsistent.

\section{The Information-Theoretic Interpretation}
The salient difference between classical and quantum mechanics is the noncommutativity of the algebra of observables of a quantum system, equivalently the non-Booleanity of the algebra of two-valued observables representing properties (for example, the property that the energy of the system lies in a certain range of values, with the two eigenvalues representing `yes' or `no'), or propositions (the proposition asserting that the value of the energy lies in this range, with the two eigenvalues representing `true' or `false'). The two-valued observables of a classical system form a Boolean algebra, isomorphic to the Borel subsets of the phase space of the system. The transition from classical to quantum mechanics replaces this Boolean algebra by a family of `intertwined' Boolean algebras, to use Gleason's term \cite{Gleason}, one for each set of commuting two-valued observables, represented by projection operators in a Hilbert space. The intertwinement precludes the possibility of embedding the whole collection into one inclusive Boolean algebra, so you can't assign truth values consistently to the propositions about observable values in all these Boolean algebras. Putting it differently, there are Boolean algebras in the family of Boolean algebras of a quantum system, for example the Boolean algebras for position and momentum, or for spin components in different directions, that don't fit together into a single Boolean algebra, unlike the corresponding family for a classical system. In this non-Boolean theory, probabilities are, as von Neumann put it, `sui generis' \cite{Neumann1} and `uniquely given from the start' \cite[p. 245]{Neumann2} via Gleason's theorem, or the Born rule, as a feature of the geometry of Hilbert space, related to the angle between rays in Hilbert space representing `pure' quantum states.\footnote{This conceptual picture applies to quantum mechanics on a finite-dimensional Hilbert space. A restriction to `normal' quantum states is required for quantum mechanics formulated with respect to a general von Neumann algebra, where a generalized Gleason's theorem holds even for quantum probability functions that are not countably additive. Thanks to a reviewer for pointing this out.}

The central interpretative question for quantum mechanics as a non-Boolean theory is how we should understand these `sui generis' probabilities, since they are not probabilities of spontaneous transitions between quantum states, nor can they be interpreted as measures of ignorance about quantum properties associated with the actual values of observables prior to measurement. 

The information-theoretic interpretation is the proposal to take Hilbert space as the kinematic framework for the physics of an indeterministic universe, just as Minkowski space provides the kinematic framework for the physics of  a non-Newtonian, relativistic universe.\footnote{See Michel Janssen \cite{Janssen} for a defense of this view of special relativity contra Harvey Brown \cite{Brown}.} In special relativity,  the geometry of Minkowski space imposes spatio-temporal constraints on events to which the relativistic dynamics is required to conform. In quantum mechanics, the non-Boolean projective geometry of Hilbert space imposes objective kinematic (i.e., pre-dynamic) probabilistic constraints on correlations between events to which a quantum dynamics of matter and fields is required to conform. In this non-Boolean theory, new sorts of nonlocal probabilistic correlations are possible for `entangled' quantum states  of separated systems, where the correlated events are  intrinsically random, not merely apparently random like coin tosses (see Bub \cite[Chapter 4]{Bub}).

In \cite{Pitowsky1}, Pitowsky distinguished between a `big' measurement problem and a `small' measurement problem.  The `big' measurement problem is the problem of explaining how measurements can have definite outcomes, given the unitary dynamics of the theory. The `small' measurement problem is the problem of accounting for our familiar experience of a classical or Boolean macroworld, given the non-Boolean character of the underlying quantum event space. The `big' problem is  the problem of explaining \emph{how individual measurement outcomes come about dynamically}, i.e., how something  indefinite in the quantum state of a system before measurement can \emph{become definite} in a measurement. There is nothing analogous to this sort of transition in classical physics, where transitions are always between an initial state of affairs specified by what is and is not the case (equivalent to a 2-valued homomorphism on a Boolean algebra) to a final state of affairs with a different specification of  what is and is not the case. The `small' problem is  the problem of explaining  the \emph{emergence of an effectively classical  probability space for the macro-events we observe in a measurement}. 

The `big' problem is deflated as a pseudo-problem if we reject  two `dogmas' about quantum mechanics. The first dogma is the view that measurement in a fundamental mechanical theory should be treated as a dynamical process, so it should be possible, in principle, to give a complete dynamical analysis of how  individual measurement outcomes come about.  The second dogma is the interpretation of a quantum state as a representation of physical reality, i.e., as the `truthmaker' for propositions about the occurrence and non-occurrence of events, analogous to the ontological significance of a classical state. 

The first dogma about measurement  is an entirely reasonable demand for a fundamental theory of motion like classical mechanics. But noncomutativity or non-Booleanity makes quantum mechanics quite unlike any theory we have dealt with before in the history of physics, and  there is no reason, apart from tradition, to assume that the theory can provide   the sort of \emph{representational} explanation we are familiar with in a  theory that is commutative or  Boolean at the fundamental level. The `sui generis' quantum probabilities can't be understood as quantifying ignorance about the pre-measurement value of an observable, as in a Boolean theory, but cash out in terms of `what you'll find if you measure,' which involves considering the outcome, at the Boolean macrolevel, of manipulating a quantum system in a certain way. 

A quantum `measurement' is  not really the same sort of thing as a measurement of a physical quantity of a classical system. It involves putting a microsystem, like a photon, in a macroscopic environment, say a beamsplitter or an analyzing filter, where the photon is forced to make an intrinsically random transition recorded as one of a number of macroscopically distinct alternatives in a macroscopic device like a photon detector. The registration of the measurement outcome at the Boolean macrolevel is crucial, because it is only with respect to a suitable structure of alternative possibilities that it makes sense to talk about an event as definitely occurring or not occurring, and this structure---characterized by Boole as capturing the  `conditions of possible experience'\cite{Boole}---is a Boolean algebra.

In special relativity, Lorentz contraction is  a kinematic effect of motion in a
non-Newtonian space-time. The contraction is \emph{consistent} with a dynamical account, but such an account takes the forces involved to be Lorentz covariant, which is to say that the dynamics is assumed to have symmetries that respect Lorentz contraction as a kinematic effect of relative motion. (By contrast, in Lorentz's theory, the contraction is  explained as a dynamical effect in Newtonian space-time.) Analogously, the loss of information in a quantum measurement---Bohr's `irreducible and uncontrollable disturbance'---is  a \emph{kinematic}  effect of \emph{any} process of gaining information of the relevant sort, irrespective of the dynamical processes involved in the
measurement process. A solution to the `small' measurement problem---a dynamical explanation for the effective emergence of classicality, i.e., Booleanity, at the macrolevel---would amount to a proof that the unitary quantum dynamics is consistent with the kinematics, analogous to a proof that relativistic dynamics is consistent with the kinematics.

We argued in `Two dogmas' that the `small' measurement problem can be resolved as a consistency problem by considering the dynamics of the measurement process and the role of decoherence in the emergence of an effectively classical probability space of macro-events to which the Born probabilities refer.  (The proposal was that decoherence solves the `small' measurement problem, not the `big' measurement problem---decoherence does not provides a dynamical explanation of how an indefinite outcome in a quantum superposition becomes definite in the unitary evolution of a measurement process.) An alternative solution is suggested by Pitowsky's combinatorial treatment of macroscopic objects in quantum mechanics \cite{Pitowsky2}. Entanglement witnesses are observables that distinguish between separable and entangled states. An entanglement witness for an entangled state is an observable whose expectation value lies in a bounded interval for a separable state, but is outside this interval for the entangled state. Pitowsky showed, for a large class of entanglement witnesses,  that for composite systems where measurement of an entanglement witness requires many manipulations of individual particles, entangled states that can be distinguished from separable states become rarer and rarer as the number of particles increases, and he conjectured that this is true in general. If Pitowsky's conjecture is correct, a macrosystem in quantum mechanics can be characterized as a many-particle system for which the measure of the set of entangled states that can be distinguished from separable states tends to zero. In this sense, a macrosystem is effectively a commutative or Boolean system.

On the information-theoretic interpretation, quantum mechanics is a new sort of \emph{non-representational} theory for an indeterministic universe, in which the quantum state is a bookkeeping device for keeping track of probabilities and probabilistic correlations between intrinsically random events. Probabilities are defined with respect to a single Boolean perspective, the Boolean algebra generated by the `pointer-readings' of what Bohr referred to as the `ultimate measuring instruments,' which are `kept outside the system subject to quantum mechanical treatment' \cite[pp. 23--24]{Bohr1}:
\begin{quote}
In the system to which the quantum mechanical formalism is applied, it is of course possible to include any intermediate auxiliary agency employed in the measuring processes. \ldots The only significant point is that in each case some ultimate measuring instruments, like the scales and clocks which determine the frame of space-time coordination---on which, in the last resort, even the definition of momentum and energy quantities rest---must always be described entirely on classical lines, and consequently be kept outside the system subject to quantum mechanical treatment. 
\end{quote}

Bohr did not, of course, refer to Boolean algebras, but the concept is simply a precise way of codifying a significant aspect of what Bohr meant by a description `on classical lines' or `in classical terms' in his constant  insistence that  (his emphasis) \cite[p. 209]{Bohr2}
\begin{quote}
\emph{however far the phenomena transcend the scope of classical physical explanation, the account of all evidence must be expressed in classical terms.}
\end{quote}
 by which he meant `unambiguous language with suitable application of the terminology of classical physics'---for the simple reason, as he put it, that we need to be able `to tell others what we have done and what we have learned.' Formally speaking, the significance of `classical' here as being able `to tell others what we have done and what we have learned' is that the events in question should fit together as a Boolean algebra, so conforming to Boole's `conditions of possible experience.'

The solution to the `small' measurement problem as a consistency problem does not show that unitarity  is suppressed at a certain level of size or complexity, so that the non-Boolean possibility structure becomes Boolean and quantum becomes classical at the macrolevel. Rather,  the claim is  that the unitary dynamics of quantum mechanics is consistent with the kinematics, in the sense that treating the measurement process dynamically and ignoring certain information that is in practice inaccessible makes it extremely difficult to detect the phenomena of interference and entanglement associated with non-Booleanity. In this sense, an effectively classical or Boolean probability space of Born probabilities can be associated with our observations at the macrolevel. 

Any system, of any complexity, is fundamentally a non-Boolean quantum system and can be treated as such, in principle, which is to say that a unitary dynamical analysis can be applied to whatever level of precision you like. But we see actual events happen at the Boolean macrolevel. At the end of a chain of instruments and recording devices, some particular system, $M$, functions as the `ultimate measuring instrument' with respect to which an event corresponding to a definite measurement outcome occurs in an associated Boolean algebra, whose selection is not the outcome of a dynamical evolution described by the theory. The system $M$, or any part of $M$, can be treated quantum mechanically, but then some other system, $M^{\prime}$, treated as classical or commutative or Boolean, plays the role of the ultimate measuring instrument in any application of the theory. The outcome of a measurement is an intrinsically random event at the macro\-level, \emph{something that actually happens}, not described by the deterministic unitary dynamics, so outside the theory. Putting it differently, the  `collapse,' as a conditionalization of the quantum state, is something you put in by hand after recording the actual outcome. The physics doesn't give it to you.

As Pauli put it \cite{Pauli}:
\begin{quote}
Observation thereby takes on the character of i\emph{rrational, unique actuality} with unpredictable outcome. \ldots Contrasted with this \emph{irrational} aspect of concrete phenomena which are determined in their \emph{actuality}, there stands the \emph{rational} aspect of an abstract ordering of the possibilities of statements by means of the mathematical concept of probability and the $\psi$-function. 
\end{quote}

A representational theory proposes a primitive ontology, perhaps of particles or fields of a certain sort, and a dynamics that describes how things change over time. The non-Boolean physics of quantum mechanics does not provide a representational explanation of phenomena. Rather, a quantum mechanical explanation involves showing how a Boolean output (a measurement outcome) is obtained from a Boolean input (a state preparation) via a non-Boolean link. If a `world' in which truth and falsity, `this' rather than `that,' makes sense is Boolean, then there is no quantum `world,' as Bohr is reported to have said,\footnote{Aage Petersen \cite[p. 12]{Petersen}: `When asked whether the algorithm of quantum mechanics could be considered as somehow mirroring an underlying quantum world, Bohr would answer,  ``There is no quantum world. There is only an abstract quantum physical description. It is wrong to think that the task of physics is to find out how nature \emph{is}. Physics concerns what we can \emph{say} about nature.''} and it would be misleading to attempt to picture it. In this sense, a quantum mechanical explanation is `operational,' but this is not simply a matter of convenience or philosophical persuasion. We adopt quantum mechanics---the theoretical formalism of the non-Boolean link between Boolean input and output---for empirical reasons, the failure of classical physics to explain certain phenomena, and because there is no satisfactory representational theory of such phenomena.

Quantum mechanics on the Everett interpretation is regarded, certainly by its proponents, as a  perfectly good representational theory: it explains phenomena in terms of an underlying ontology and a dynamics that accounts for change over time. As I will show in the following section, the Everett interpretation leads to modal contradictions in scenarios that involve `encapsulated' measurements.

\section{Encapsulated Measurements and the Everett Interpretation}

According to the Born rule, the probability of finding the outcome $a$ in a measurement of an observable $A$ on a system $S$ in the state $\ket{\psi} = \sum_{a}\langle a|\psi\rangle\ket{a}$ is:
\begin{equation}
p_{\psi}(a) = \trace{P_{a}P_{\psi}} = |\langle a|\psi\rangle|^{2}
\end{equation}
where $P_{a}$ is the projection operator onto the eigenstate $\ket{a}$ and $P_{\psi}$ is the projection operator onto the quantum state. After the measurement, the state is updated to
\begin{equation}
\ket{\psi} \longrightarrow \ket{a}
\end{equation}
On the information-theoretic interpretation, the state update is understood as conditionalization (with necessary loss of information) on the measurement outcome. 

For a sequence of measurements, perhaps by two different observers, of $A$ followed by $B$ on the same system $S$ initially in the state $\ket{\psi}$, where $A$ and $B$ need not commute, the conditional probability of the outcome $b$ of $B$ given the outcome $a$ of $A$ is 
\begin{equation}
p_{\psi}(b|a) = \frac{p_{\psi}(a,b)}{\sum_{b} p_{\psi}(a,b)} = \frac{p_{\psi}(a,b)}{p_{\psi}(a)} = \trace{P_{b}P_{a}} = |\langle b|a\rangle|^{2}
\end{equation}
with $p_{\psi}(a,b) = |\langle a|\psi\rangle|^{2} \cdot |\langle b|a\rangle|^{2} = p_{\psi}(a)\cdot |\langle b|a\rangle|^{2} $.

According to Everett's relative state interpretation, a measurement is a unitary transformation, or equivalently an isometry, $V$, that correlates the pointer-reading state of the measuring instrument and the associated memory state of an observer with the state of the observed system. (In the following, I'll simply refer to the memory state, with the understanding that this includes the measuring instrument and recording device and any other systems involved in the measurement.) In the case of a projective measurement by an observer $O$, the transformation $\mathcal{H}_{S} \stackrel{V}{\longrightarrow} \mathcal{H}_{S}\otimes\mathcal{H}_{O}$ maps $\ket{a}_{S}$ onto $\ket{a}_{S}\otimes\ket{A_{a}}_{O}$, where $\ket{A_{a}}_{O}$ is the memory state correlated with the eigenstate $\ket{a}_{S}$, so 
\begin{equation}
\ket{\psi} \stackrel{V}{\longrightarrow} \ket{\Psi} = \sum_{a}\langle a|\psi\rangle\ket{a}_{S} \otimes \ket{A_{a}}_{O}
\end{equation}

Baumann and Wolf \cite{Baumann+} formulate the Born rule for the memory state $\ket{A_{a}}_{O}$ of the observer $O$  having observed $a$ in the relative state interpretation, where there is no commitment to the existence of a physical record of the measurement outcome as classical information. The probability of the observer $O$ observing $a$ in this sense is 
\begin{equation}
 q_{\psi}(a) = \trace{\mathds{1} \otimes P_{A_{a}}\cdot V P_{\psi}V^{\dagger}}
 \end{equation}
where $P_{A_{a}}$ is the projection operator onto the memory state $\ket{A_{a}}_{O}$. This is equivalent to the probability $p_{\psi}(a)$ in the standard theory. 

This equivalence can also be interpreted as showing  the movability of the notorious Heisenberg `cut' between the observed system and the composite system doing the observing, if the observing system is treated as a classical or Boolean system. The cut can be placed between the system and the measuring instrument plus observer, so that only the system is treated quantum mechanically, or between the measuring instrument and observer, so that the system plus instrument is treated as a composite quantum system. Similar shifts are possible if the measuring instrument is subdivided into several component systems, e.g., if a recording device is regarded as a separate component system, or if the observer is subdivided into component systems involved in the registration of a measurement outcome.

On the relative state interpretation, the probability that $S$ is in the state $\ket{a'}$ but the observer sees $a$ is
\begin{equation}
q_{\psi}(\ket{a'},a) = \trace{P_{a'} \otimes P_{A_{a}}\cdot V P_{\psi}V^{\dagger}} = \delta_{a',a}
\end{equation}
So  the probability that $S$ is in the state $\ket{a}$ after $O$ observes $a$  is 1, as in the standard theory according to the state update rule.

For a sequence of measurements of observables $A$ and $B$ by two observers, $O_{1}$ and $O_{2}$, on the same system $S$, the conditional probability for the outcome $b$ given the outcome $a$ is
\begin{equation}
q_{\psi}(b|a)  = = \frac{q_{\psi}(a,b)}{\sum_{b} q_{\psi}(a,b)} =  \frac{\trace{\mathds{1}\otimes P_{A_{a}}\otimes P_{B_{b}}\cdot V_{O_{2}}V_{O_{1}}P_{\psi}V_{O_{1}}^{\dagger}V_{O_{2}}^{\dagger} }}{\sum_{b}\trace{\mathds{1}\otimes P_{A_{a}}\otimes P_{B_{b}}\cdot V_{O_{2}}V_{O_{1}}P_{\psi}V_{O_{1}}^{\dagger}V_{O_{2}}^{\dagger} }} \label{eq:relstatecond}
\end{equation}
which turns out to be the same as the conditional probability given by the state update rule: $q_{\psi}(b|a) =p_{\psi}(b|a)$.

So Everett's relative state interpretation and the information-theoretic interpretation make the same predictions, both for the probability of an outcome of an $A$-measurement on a system $S$, and for the conditional probability of an outcome of a $B$-measurement, given an outcome of a prior $A$-measurement  on the same system $S$ by two observers, $O_{1}$ and $O_{2}$. 

As Baumann and Wolf show, this is no longer the case for conditional probabilities involving encapsulated measurements at different levels of observation, where an observer, $O$ measures an observable of a system $S$ and a super-observer measures an arbitrary observable of the joint system $S+O$. In other words, the movability of the cut is restricted to measurements that are not encapsulated. Note that both the observer and the super-observer could be quantum automata, so the analysis is not restricted to observers as conscious agents.

Suppose an observer $O$ measures an observable with eigenstates $\ket{a}_{S}$ on a system $S$ in a state $\ket{\psi}_{S} = \sum_{a}\langle a|\psi\rangle\ket{a}_{S}$, and a super-observer $SO$ then measures an observable with eigenstates $\ket{b}_{S+O}$ in the Hilbert space of the composite system $S+O$.

On the information-theoretic interpretation of the standard theory, the state of $S+O$ after the measurement by $O$ is $\ket{a \otimes A_{a}}_{S+O} = \ket{a}_{S}\otimes\ket{A_{a}}_{O}$, where $\ket{A_{a}}_{O}$ is the memory state of $O$ correlated with the outcome $a$. The probability of the super-observer finding the outcome $b$ given that the observer $O$ obtained the outcome $a$ is then
\begin{equation}
p_{\psi}(b|a) = |\langle b|a \otimes A_{a}\rangle_{S+O}|^{2} \label{eqn:colltheorycond}
\end{equation}

On the relative state interpretation,  after the unitary evolutions associated with the observer $O$'s measurement of the  $S$-observable with eigenstates $\ket{a}$, and the super-observer $SO$'s measurement of the $(S+O)$-observable with eigenstates $\ket{b}$, the state of the composite system $S+O+SO$ is
\begin{equation}
\ket{\Psi} = \sum_{a,b}\langle a|\psi\rangle\langle b|a\otimes A_{a}\rangle |b\rangle_{S+O} |B_{b}\rangle_{SO} \label{uncollapsed}
\end{equation}
where $\ket{B_{b}}_{O}$ is the memory state of $SO$ correlated with the outcome $b$. The probability of the super-observer $SO$ finding the outcome $b$ given that the observer $O$ obtained the outcome $a$ is then $q_{\psi}(b|a)$ as given by (\ref{eq:relstatecond}), with $\ket{\Psi}\bra{\Psi} = V_{O_{2}}V_{O_{1}}P_{\psi}V_{O_{1}}^{\dagger}V_{O_{2}}^{\dagger}$. The numerator of this expression, after taking the trace over the $(S+O)$-space followed by the trace over the $O$-space of  the projection onto $P_{B_{b}}$ followed by the projection onto $P_{A_{a}}$,  becomes
\begin{equation}
p_{\psi}(b|a)\sum_{a',a''}\langle a'|\psi\rangle\langle\psi |a''\rangle\langle b|a'\otimes A_{a'}\rangle\langle a''\otimes A_{a''}|b\rangle
\end{equation}
so 
\begin{equation}
q_{\psi}(b|a) = \frac{p_{\psi}(b|a)\sum_{a',a''}\langle a'|\psi\rangle\langle\psi |a''\rangle\langle b|a'\otimes A_{a'}\rangle\langle a''\otimes A_{a''}|b\rangle}{\sum_{b}\trace{\mathds{1}\otimes P_{A_{a}}\otimes P_{B_{b}}\cdot  | \Psi\rangle\langle\Psi |}}
\end{equation}
Baumann and Wolf point out that this is equal to $p_{\psi}(b|a)$ in (\ref{eqn:colltheorycond}) only if $\ket{b}_{S+O}= \ket{a}_{S}\otimes \ket{A_{a}}_{O}$ for all $b$, i.e., if $\ket{b}_{S+O}$ is a product state of an $S$-state $\ket{a}_{S}$ and an $O$-state $\ket{A_{a}}_{O}$, which is not the case for general encapsulated measurements. 

Here $O$'s measurement outcome states $\ket{A_{a}}_{O}$ are understood as relative to  states of $SO$, and there needn't be a record of $O$'s measurement outcome as classical or Boolean information. If there is a classical record of the outcome then, as Baumann and Wolf show, $SO$'s conditional probability is the same as the conditional probability in the standard theory.

To see that the relative state interpretation leads to a modal contradiction, consider the Frauchiger-Renner modification of the `Wigner's Friend' thought experiment \cite{Frauchiger+}. Frauchiger and Renner add a second Friend ($\overline{F}$) and a second Wigner ($\overline{W}$). Friend $\overline{F}$  measures an observable with eigenvalues `heads' or `tails' and eigenstates $\ket{h}, \ket{t}$ on a system $R$ in the state $\frac{1}{\sqrt{3}}\ket{h} + \sqrt{\frac{2}{3}}\ket{t}$. One could say that $\overline{F}$ `tosses a biased quantum coin' with probabilities 1/3 for heads and 2/3 for tails. She prepares a qubit $S$ in the state $\ket{\downarrow}$ if the outcome is $h$, or in the state  $\ket{\rightarrow} = \frac{1}{\sqrt{2}}(\ket{\downarrow} + \ket{\uparrow})$ if the outcome is $t$, and sends $S$ to  $F$. When  $F$ receives $S$, he measures an $S$-observable with eigenstates $\ket{\downarrow}, \ket{\uparrow}$ and corresponding eigenvalues $-\frac{1}{2}, \frac{1}{2}$. The Friends $\overline{F}$ and $F$ are in two laboratories, $\overline{L}$ and $L$, which are assumed to be completely isolated from each other, except briefly when $\overline{F}$ sends the qubit $S$ to $F$.  

Now suppose that  $\overline{W}$ is able to measure an $\overline{L}$-observable with eigenstates 
\begin{eqnarray}
\ket{\mbox{fail}}_{\overline{L}} & = &  \frac{1}{\sqrt{2}}(\ket{h}_{\overline{L}} + \ket{t}_{\overline{L}})  \nonumber \\
\ket{\mbox{ok}}_{\overline{L}} & = & \frac{1}{\sqrt{2}}(\ket{h}_{\overline{L}} - \ket{t}_{\overline{L}}) \nonumber
\end{eqnarray}
and $W$ is able to measure an $L$-observable   with eigenstates 
\begin{eqnarray}
\ket{\mbox{fail}}_{L} & = &  \frac{1}{\sqrt{2}}(\ket{-\frac{1}{2}}_{L} + \ket{\frac{1}{2}}_{L}) \nonumber \\
\ket{\mbox{ok}}_{L} & =  & \frac{1}{\sqrt{2}}(\ket{-\frac{1}{2}}_{L} - \ket{\frac{1}{2}}_{L}) \nonumber
\end{eqnarray} 

Suppose $W$ measures first and we stop the experiment at that point. On the relative state interpretation,  $\overline{F}$'s memory and all the systems in $\overline{F}$'s laboratory involved in the measurement of $R$ become entangled by the unitary transformation, and similarly $F$'s memory and all the systems in $F$'s laboratory involved in the measurement of $S$ become entangled. If $R$ is in the state $\ket{h}_{R}$ and  $\overline{F}$ measures $R$, the entire laboratory $\overline{L}$ evolves to the state $\ket{h}_{\overline{L}} = \ket{h}_{R}\ket{h}_{\overline{F}}$, where  $\ket{h}_{\overline{F}}$ represents the  state of $\overline{F}$'s memory plus measuring and recording device plus all the systems in $\overline{F}$'s lab connected to the measuring and recording device after the registration of outcome $h$. Similarly for the state $\ket{t}_{\overline{L}}$, and for the states $\ket{-\frac{1}{2}}_{L} = \ket{-\frac{1}{2}}_{S}\ket{-\frac{1}{2}}_{F}$ and $\ket{\frac{1}{2}}_{L} = \ket{\frac{1}{2}}_{S}\ket{\frac{1}{2}}_{F}$ of $F$'s lab $L$ with respect to the registration of the outcomes $\pm\frac{1}{2}$ of the spin measurement. So after the evolutions associated with the measurements by $\overline{F}$ and $F$, the state of the two laboratories is
\begin{eqnarray}
\ket{\Psi} & = & \frac{1}{\sqrt{3}}\ket{h}_{\overline{L}}\ket{-\frac{1}{2}}_{L} + \sqrt{\frac{2}{3}} \ket{t}_{\overline{L}} \frac{ \ket{-\frac{1}{2}}_{L}+ \ket{\frac{1}{2}}_{L}}{\sqrt{2}} \label{eqn:entangled1}   \\
& = & \frac{1}{\sqrt{3}}\ket{h}_{\overline{L}}\ket{-\frac{1}{2}}_{L} + \sqrt{\frac{2}{3}}\ket{t}_{\overline{L}}\ket{\mbox{fail}}_{L} \label{eqn:entangled2} 
\end{eqnarray}
According to the state (\ref{eqn:entangled2}), the probability is zero that $W$ gets the outcome `ok' for his measurement, given that $\overline{F}$ obtained the outcome $t$ for her measurement.

Now suppose that $\overline{W}$ and $W$ both measure, and $\overline{W}$ measures before $W$, as in the Frauchiger-Renner scenario. What is the probability that $W$ will get `ok' at the later time given that $\overline{F}$ obtained `tails' at the earlier time? The global entangled state at any time simply expresses a correlation between measurements---the time order of the measurements is irrelevant. The two measurements by $W$ and $\overline{W}$ could be spacelike connected, and in that case the order in which the measurements are carried out clearly can't make a different to the probability.\footnote{Thanks to Renato Renner for pointing this out.}

Since observers in differently moving reference frames agree about which events occur, even if they disagree about the order of events,  an event that has zero probability in some reference frame cannot occur for any observer in any reference frame.\footnote{In \cite{BubStairs}, Allen Stairs and I proposed this as a consistency condition to avoid potential contradictions in quantum interactions with closed timelike curves.} In  a reference frame in which $W$ measures before $\overline{W}$ the probability is zero that $W$ gets `ok' if $\overline{F}$ gets `tails,' because this must be the same as the probability if we take $\overline{W}$ out of the picture. It follows that if $\overline{F}$ gets `tails,' the `ok' measurement outcome event cannot occur in a reference frame in which $\overline{W}$ measures before $W$. 

Think of it from $\overline{F}$'s perspective. If she is certain that she found `tails,' she can predict with certainty that $W$'s later measurement (timelike connected to her `tails' event) will not result in the outcome `ok.' She wouldn't (shouldn't!) change her prediction because of the possible occurrence of an event spacelike connected to 
$W$'s measurement---not if she wants to be consistent with special relativity. $W$ finding `ok' is a zero probability event in the absolute future of $\overline{F}$'s prediction that cannot occur for $\overline{F}$ given the `tails' outcome of her $R$-measurement, and hence cannot occur in any reference frame. The fact that $\overline{W}$'s measurement, an event spacelike connected to $W$'s measurement, might occur after her prediction doesn't support altering the prediction, even though it would make $\overline{F}$'s memory of the earlier event (and any trace of the earlier event in $\overline{F}$'s laboratory $L$) indefinite, so no observer could be in a position to check whether or not $\overline{F}$ obtained the earlier outcome $t$ when $W$ gets `ok.'

The state (\ref{eqn:entangled2}) can be expressed as
\begin{equation}
\ket{\Psi} = \frac{1}{\sqrt{12}}\ket{\mbox{ok}}_{\overline{L}}\ket{\mbox{ok}}_{L} - \frac{1}{\sqrt{12}}\ket{\mbox{ok}}_{\overline{L}}\ket{\mbox{fail}}_{L} + \frac{1}{\sqrt{12}}\ket{\mbox{fail}}_{\overline{L}}\ket{\mbox{ok}}_{L} +\frac{\sqrt{3}}{2}\ket{\mbox{fail}}_{\overline{L}}\ket{\mbox{fail}}_{L} \label{eqn:entangled3} 
\end{equation}
It follows that the probability that both Wigners obtain the outcome `ok' is 1/12. Since the two Wigners measure commuting observables on separate systems, $\overline{W} $ can communicate the outcome `ok' of her measurement to $W$, and her prediction that she is certain, given the outcome `ok,'  that $W$ will obtain `fail,' without `collapsing' the global entangled state. Then in a round in which $W$ obtains the outcome `ok' for his measurement and so is certain that the outcome is `ok,' he is also certain that the outcome of his measurement is not `ok.' 

Frauchiger and Renner derive this modal contradiction on the basis of three assumptions,  Q, S, and C. Assumption Q (for `quantum') says that an agent can `be certain' of the value of an observable at time $t$ if the quantum state assigns probability 1 to this value after a measurement that is completed at time $t$, and the agent has established that the system is in this state. An agent can also `be certain' that the outcome does not occur if the probability is zero. One should also include under Q the assumption that measurements are quantum interactions that transform the global quantum state unitarily. Assumption S (for `single value')  says, with respect to `being certain,' that an agent can't `be certain' that the value of an observable is equal to $v$ at time $t$ and also `be certain' that the value of this observable is not equal to $v$ at time $t$.\footnote{The standard axiom system for the modal logic of the operator `I am certain that' includes the axiom `if I am certain that $A$, then it is not the case that I am certain that not-$A$'  (assuming the accessibility relation is serial and does not have any dead-ends). Thanks to Eric Pacuit for pointing this out.}  Finally, assumption C (for `consistency')  is a transitivity condition for `being certain': if an agent `is certain' that another agent, reasoning with the same theory, `is certain' of the value of an observable at time $t$, then the first agent can also `be certain' of this value at time $t$.

As used in the Frauchiger-Renner argument, `being certain' is a technical term entirely specified by the three assumptions. In other words, `being certain' can mean whatever you like, provided that the only inferences involving the term are those sanctioned by the three assumptions. In particular, it is not part of the Frauchiger-Renner argument that if an agent `is certain' that the value of an observable is $v$ at time $t$, then the value of the observable is indeed $v$ at time $t$---the observable might have multiple values, as in the Everett interpretation. So it does not follow that the proposition `the value of the observable is $v$ at time $t'$' is true, or that the system has the property corresponding to the value $v$ of the observable at time $t$. Also, it is not part of the Frauchiger-Renner argument that if an agent `is certain' of the value of an observable, then the agent knows the value, in any sense of `know' that entails the truth of what the agent knows. 

The argument involves measurements, and inferences by the agents via the notion of `being certain,' that can all be modeled as unitary transformations to the agents' memories. So picture the agents as quantum automata, evolving unitarily, where these unitary evolutions result in changes to the global quantum state of the two friends and the two Wigners that reflect changes in the agents' memories. What Frauchiger and Renner show is that these assumptions lead to memory changes that end up in a modal contradiction. 

Obviously, the relative state interpretation is inconsistent with the assumption S in the sense that different branches of the quantum state can be associated with  different values of an observable after a measurement. So, with respect to a particular branch, an agent can be certain that an observable has a certain value and, with respect to a different branch, an agent associated with that branch can be certain that the observable has a different value. This is not a problem. Rather, the problem for the relative state interpretation is that there is \emph{a branch}  of the global quantum state after $W$'s measurement, which has a non-zero probability  of 1/12, with a memory entry that is inconsistent with S: $W$ is  certain that  the outcome of his measurement is `fail' and also certain that the outcome is `ok.'

Renner gives a full description of the state of the global system as it evolves unitarily with the measurements and inferential steps at every stage of the experiment, which is repeated over many rounds until $\overline{W}$ and $W$ both obtain the outcome 'ok'  \cite{Renner}. The measurements are assumed to be performed by the agents $\overline{F}, F, \overline{W}, W$ in sequence beginning at times 00, 10, 20, and 30 (and ending at times 01, 11, 21, and 31). 

So, for example, in round $n$ after $F$'s measurement the global entangled state is $ \ket{\Psi}^{n:11}$ (cf. (\ref{eqn:entangled1})):
\begin{eqnarray}
&&
 \frac{1}{\sqrt{3}}\ket{h}_{\overline{R}}\ket{\scriptsize{h, \mbox{no conclusion}}}_{\overline{F}}\ket{\downarrow}_{S}\ket{\scriptsize{z=-\frac{1}{2}}}_{F}  \nonumber \\
&&   + \sqrt{\frac{1}{3}}\ket{t}_{\overline{R}}\ket{\scriptsize{t, \mbox{so I am certain that $w$ = fail at  n:31}}}_{\overline{F}}\ket{\downarrow}_{S}\ket{\scriptsize{z=-\frac{1}{2}}}_{F}  \nonumber \\
 &&  +  \sqrt{\frac{1}{3}}\ket{t}_{\overline{R}}\ket{\scriptsize{t, \mbox{so I am certain that $w$ = fail at  n:31}}}_{\overline{F}}\ket{\uparrow}_{S}\ket{\scriptsize{z=\frac{1}{2}}}_{F} 
 \end{eqnarray}
After the measurements by $\overline{W}$ and $W$ with outcomes $\bar{w}$ and $w$, the global entangled state  is $ \ket{\Psi}^{n:31}$:
\begin{eqnarray}
&& \sqrt{\frac{1}{12}}\ket{\mbox{\scriptsize{ok}}}_{\overline{L}}\ket{\scriptsize{\mbox{ok, so I am certain that $w$ = fail}}}_{\overline{W}}\ket{\mbox{ok}}_{L}\ket{\scriptsize{\mbox{I am certain that $w$ = fail; I observe $w$ = ok!}}}_{W} \nonumber \\
&&   - \sqrt{\frac{1}{12}}\ket{\mbox{\scriptsize{ok}}}_{\overline{L}}\ket{\scriptsize{\mbox{ok, so I am certain that $w$ = fail}}}_{\overline{W}}\ket{\mbox{fail}}_{L}\ket{\scriptsize{\mbox{I am certain that $w$ = fail; I observe $w$ = fail}}}_{W}  \nonumber \\
 && + \sqrt{\frac{1}{12}}\ket{\mbox{\scriptsize{fail}}}_{\overline{L}}\ket{\scriptsize{\mbox{fail, no conclusion}}}_{\overline{W}}\ket{\mbox{ok}}_{L}\ket{\scriptsize{\mbox{no conclusion previously; I observe $w$ = ok}}}_{W}  \nonumber \\
 && + \frac{\sqrt{3}}{2}\ket{\mbox{\scriptsize{fail}}}_{\overline{L}}\ket{\scriptsize{\mbox{fail, no conclusion}}}_{\overline{W}}\ket{\mbox{fail}}_{L}\ket{\scriptsize{\mbox{no conclusion previously; I observe $w$ = fail}}}_{W} 
 \end{eqnarray}
The inconsistency with S is apparent on the first branch.

\section{A Clarification}

Baumann and Wolf argue that standard quantum mechanics with the `collapse' rule for measurements, and quantum mechanics on Everett's relative state interpretation, are really two different \emph{theories}, not different \emph{interpretations} of the same theory. For the Frauchiger-Renner scenario, they derive the conditional probability
\begin{equation}
p_{\Psi}(\mbox{ok}_{W}|t_{\overline{F}}) = 0
\end{equation}
 for the standard theory  by supposing that $\overline{F}$ updates the quantum state on the basis of the outcome of her measurement of $R$ and corresponding state preparation of $S$, but $F$'s measurement is described as a unitary transformation from $\overline{F}$'s perspective, so the state of the two labs after the measurements by $\overline{F}$ (assuming $\overline{F}$ found `tails') and $F$ is $\ket{t}_{\overline{L}}(\ket{-\frac{1}{2}}_{L} + \ket{+\frac{1}{2}}_{L}) = \ket{t}_{\overline{L}}\ket{\mbox{fail}}_{L}$. 
For the relative state theory,  they derive the conditional probability
\begin{equation}
q_{\Psi}(\mbox{ok}_{W}|t_{\overline{F}}) = 1/6
\end{equation}
This would seem to be in conflict with the claim in the previous section that the probability is zero that $W$ gets the outcome `ok' for his measurement, given that $\overline{F}$ obtained the outcome `tails' for her measurement, whether or not $\overline{W}$ measures before $W$.

To understand the meaning of the probability $q_{\Psi}(\mbox{ok}_{W}|t_{\overline{F}})$ as Baumann and Wolf define it, note that after the unitary evolution associated with $\overline{W}$'s measurement on the lab $\overline{L}$, the state of the two laboratories and $\overline{W}$ is (from (\ref{eqn:entangled3}))
\begin{eqnarray}
\Phi & = & \frac{1}{\sqrt{12}}\ket{\mbox{ok}}_{\overline{L}}\ket{\mbox{ok}}_{\overline{W}}\ket{\mbox{ok}}_{L} - \frac{1}{\sqrt{12}}\ket{\mbox{ok}}_{\overline{L}}\ket{\mbox{ok}}_{\overline{W}}\ket{\mbox{fail}}_{L} \nonumber \\ &&
+ \frac{1}{\sqrt{12}}\ket{\mbox{fail}}_{\overline{L}}\ket{\mbox{fail}}_{\overline{W}}\ket{\mbox{ok}}_{L} +\frac{\sqrt{3}}{2}\ket{\mbox{fail}}_{\overline{L}}\ket{\mbox{fail}}_{\overline{W}}\ket{\mbox{fail}}_{L} \label{eqn:original}
\end{eqnarray}
which can be expressed as
\begin{eqnarray}
 \Phi & = & \frac{1}{\sqrt{2}}\ket{h}_{\overline{L}}\Big[\sqrt{\frac{5}{6}}\Big(\frac{3}{\sqrt{10}}\ket{\mbox{fail}}_{\overline{W}} - \frac{1}{\sqrt{10}}\ket{\mbox{ok}}_{\overline{W}}\Big)\ket{\mbox{fail}}_{L} \nonumber \\ && 
+ \frac{1}{\sqrt{6}}\Big(\frac{1}{\sqrt{2}}\ket{\mbox{fail}}_{\overline{W}}+ \frac{1}{\sqrt{2}}\ket{\mbox{ok}}_{\overline{W}}\Big)\ket{\mbox{ok}}_{L}\Big] \nonumber \\ &&
\frac{1}{\sqrt{2}}\ket{t}_{\overline{L}}\Big[\sqrt{\frac{5}{6}}\Big(\frac{3}{\sqrt{10}}\ket{\mbox{fail}}_{\overline{W}} + \frac{1}{\sqrt{10}}\ket{\mbox{ok}}_{\overline{W}}\Big)\ket{\mbox{fail}}_{L} \nonumber \\ &&
+ \frac{1}{\sqrt{6}}\Big(\frac{1}{\sqrt{2}}\ket{\mbox{fail}}_{\overline{W}}- \frac{1}{\sqrt{2}}\ket{\mbox{ok}}_{\overline{W}}\Big)\ket{\mbox{ok}}_{L}\Big] \label{eqn:healey}
\end{eqnarray}
Equation (\ref{eqn:healey}) is equivalent to (19) in Healey \cite{Healey}.

It follows from (\ref{eqn:healey}) that the joint probability $q_{\Phi}(\mbox{ok}_{W}, t_{\overline{L}})$ is defined after $\overline{W}$'s measurement and  equals 1/12. Since $q_{\Phi}(\mbox{ok}_{W}, t_{\overline{L}}) + q_{\Phi}(\mbox{fail}_{W}, t_{\overline{L}}) = 1/12 + 5/12 = 1/2$, 
\begin{equation}
q_{\Phi}(\mbox{ok}_{W}|t_{\overline{L}}) = \frac{q_{\Phi}(\mbox{ok}_{W}, t_{\overline{L}})}{q_{\Phi}(\mbox{ok}_{W}, t_{\overline{L}}) + q_{\Phi}(\mbox{fail}_{W}, t_{\overline{L}}) }= 1/6
\end{equation}
 
The conditional probability $q_{\Phi}(\mbox{ok}_{W}|t_{\overline{L}})$ is derived from the \emph{jointly observable statistics} at the latest time in the unitary evolution, so at the time immediately after $W$'s measurement, following $\overline{W}$'s prior measurement.\footnote{Thanks to Veronika Baumann for clarifying this.} After $\overline{W}$'s measurement of the observable with eigenvalues `ok,' `fail,', the $\overline{L}$-observable with eigenvalues `heads,' `tails' is indefinite. The probability $q_{\Phi}(\mbox{ok}_{W}|t_{\overline{L}})$ refers to a situation in which a super-observer $\overline{\overline{W}}$ measures an observable with eigenvalues `heads' or 'tails' on $\overline{L}$ and notes the fraction of cases where $W$ gets `ok' to those in which this measurement results in the outcome `tails.'  The `tails' value in this measurement is randomly related to the `tails' outcome of $F$'s previous measurement before $\overline{W}$'s intervention, so the probability  $q_{\Phi}(\mbox{ok}_{W}|t_{\overline{L}})$, which Baumann and Wolf identify with $q_{\Phi}(\mbox{ok}_{W}|t_{\overline{F}})$, is not relevant to $\overline{F}$'s prediction that $W$ will find `fail' in the cases where she finds `tails.' Certainly, given the disruptive nature of  $\overline{W}$'s measurement, it is not the case that $\overline{\overline{W}}$ will find `tails' after $\overline{W}$'s measurement if and only if $\overline{F}$'s measurement resulted in the outcome `tails'  at the earlier time before $\overline{W}$'s measurement.\footnote{Thanks to Renato Renner for this observation.} 

This is not a critique of Baumann and Wolf, who make no such claim. The purpose of this section is simply to clarify the difference between  the conditional probability $q_{\Phi}(\mbox{ok}_{W}|t_{\overline{F}})$, as Baumann and Wolf define it, and $\overline{F}$'s conditional prediction that $W$ will find `fail' in the cases where she finds `tails.'

\section{Concluding Remarks}

Frauchiger and Renner present their argument as demonstrating that  any interpretation of quantum mechanics that accepts assumptions Q, S, and C is inconsistent, and they point out which assumptions are rejected by specific interpretations. For the relative state interpretation and the many-worlds interpretation, they say that these interpretations conflict with assumption S, because 
\begin{quote}
any quantum measurement results in a branching into different `worlds,' in each of which one of the possible measurement outcomes occurs.
\end{quote}
But this is not the issue. Rather, Frauchiger and Renner show something much more devastating: that for a particular scenario with encapsulated measurements involving multiple agents, there is a branch of the global quantum state with a contradictory memory entry.

An interpretation of quantum mechanics is a proposal to reformulate quantum mechanics as a Boolean theory in the representational sense, either by introducing hidden variables, or by proposing that every possible outcome occurs in a measurement, or in some other way. An implicit assumption of the Frauchiger-Renner argument is that quantum mechanics is understood as a representational theory, in the minimal sense that observers can be represented as physical systems, with the possibility that observers can observe other observers.  What the Frauchiger-Renner argument really shows is that quantum mechanics can't be interpreted as a representational theory at all.

\section{Acknowledgements}
Thanks to Veronika Baumann, Michael Dascal, Allen Stairs, and Tony Sudbery for critical comments on earlier drafts.

\end{document}